# Measuring $^{26}$Al and $^{60}$Fe in the Galaxy

Roland Diehl

*Max Planck Institut für extraterrestrische Physik, D-85748 Garching, Germany*

**Abstract.** With the SPI high-resolution spectrometer on INTEGRAL, new results have been obtained for long-lived radioactive $^{26}$Al and $^{60}$Fe in our Galaxy: $^{26}$Al sources apparently share the pattern of Galactic rotation in the inner Galaxy, and thus allow to estimate a total mass of $^{26}$Al in the Galaxy of 2.8 $M_\odot$ from the measured flux. $^{60}$Fe production in massive stars is constrained by recent gamma-ray detections, and appears to be lower than predicted by standard models. We show the broader implications of these findings both for the study of our Galaxy, and for nucleosynthesis in massive stars.

## OBSERVATIONS WITH INTEGRAL

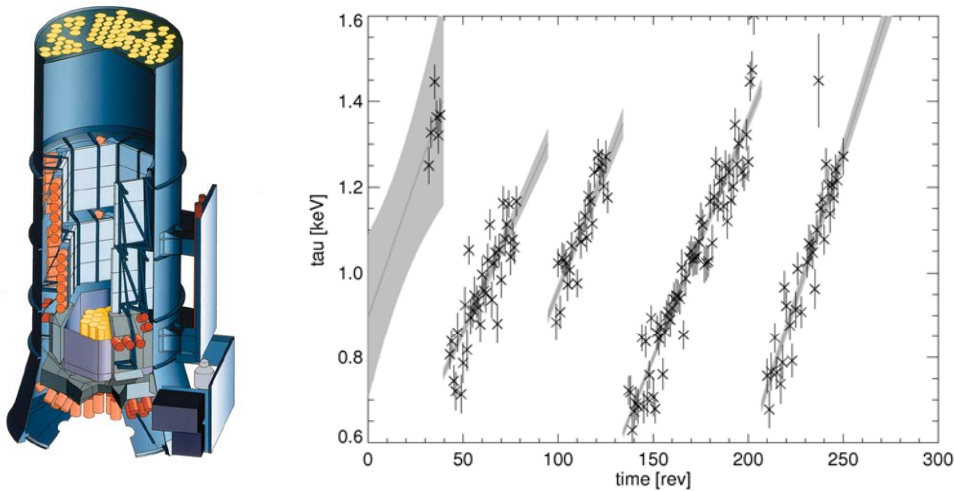

**FIGURE 1.** The SPI instrument[4,5] is built around a 19-element camera of Ge detectors; incident gamma-rays will cast a characteristic shadow onto this camera due to the coded mask, which allows to discriminate sources against instrumental background (left). Cosmic-ray bombardement in space destroys the charge-collection properties of the Ge detectors. Periodic annealing cures these defects, such that the high spectral resolution can be maintained over years. Shown is the degradation parameter τ versus time in units of INTEGRAL's 3-day orbits[3] (right)

Gamma-rays from radioactive by-products of nucleosynthesis ejecta provide a rather direct measurement of cosmic nucleosynthesis, and, in particular in the cases of long-lived isotopes such as $^{26}$Al and $^{60}$Fe, provide insights into turbulent phases of the interstellar medium around massive stars. Launches of high-resolution solid state detectors (RHESSI [2], INTEGRAL/SPI [3,4,5]) into space have added a new quality to this field (Fig.1): Fine spectral resolution allows to better identify these nuclear lines above background, and spectroscopy can constrain the kinematics of the isotopes in the gamma-ray emission region, which is astrophysically important. One demonstrated success of such recent gamma-ray spectroscopy are the signatures of beaming in the accelerated-particle flow of solar flares [6,7].

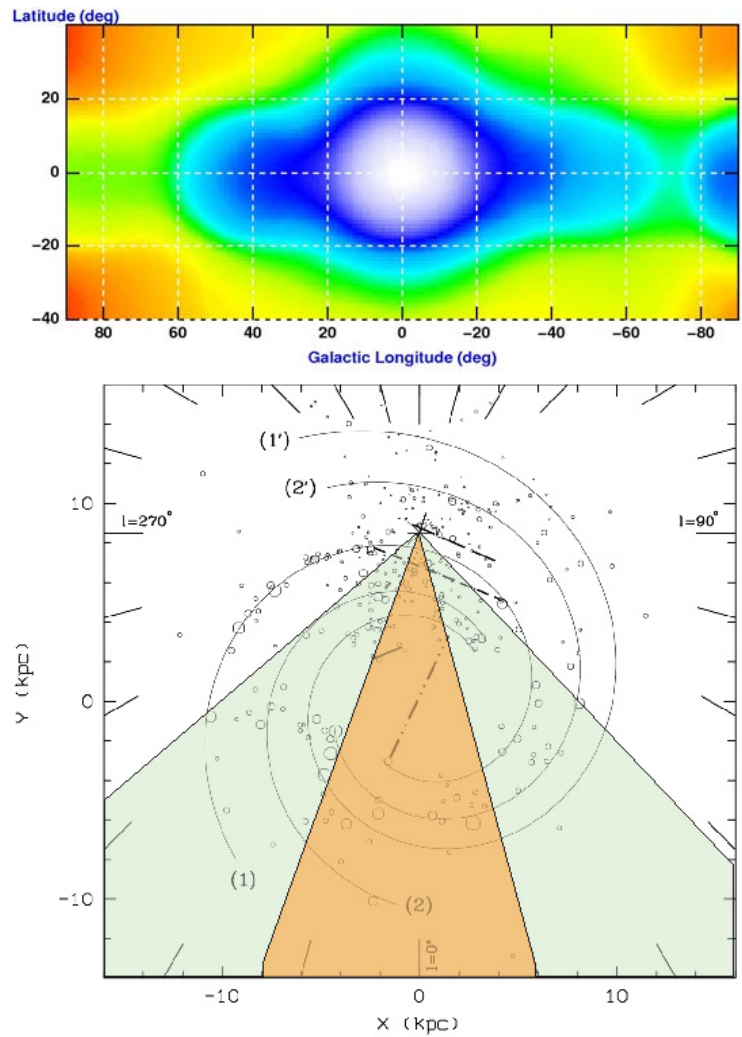

**FIGURE 2:** The INTEGRAL survey of the inner Galaxy concentrates on a longitude region between about –40° to +50° (above, and outer shaded cone in the bottom figure). In this area, the bright ridge of inner-Galaxy gamma-ray emission from $^{26}$Al covering longitudes –30° to +35° (inner shaded cone in lower figure) is well exposed. Above figure shows the star forming complexes as we know them in the Galaxy [1]. It is evident that $^{26}$Al emission explores the inner Galaxy regions, where identifications of star-forming complexes is difficult.

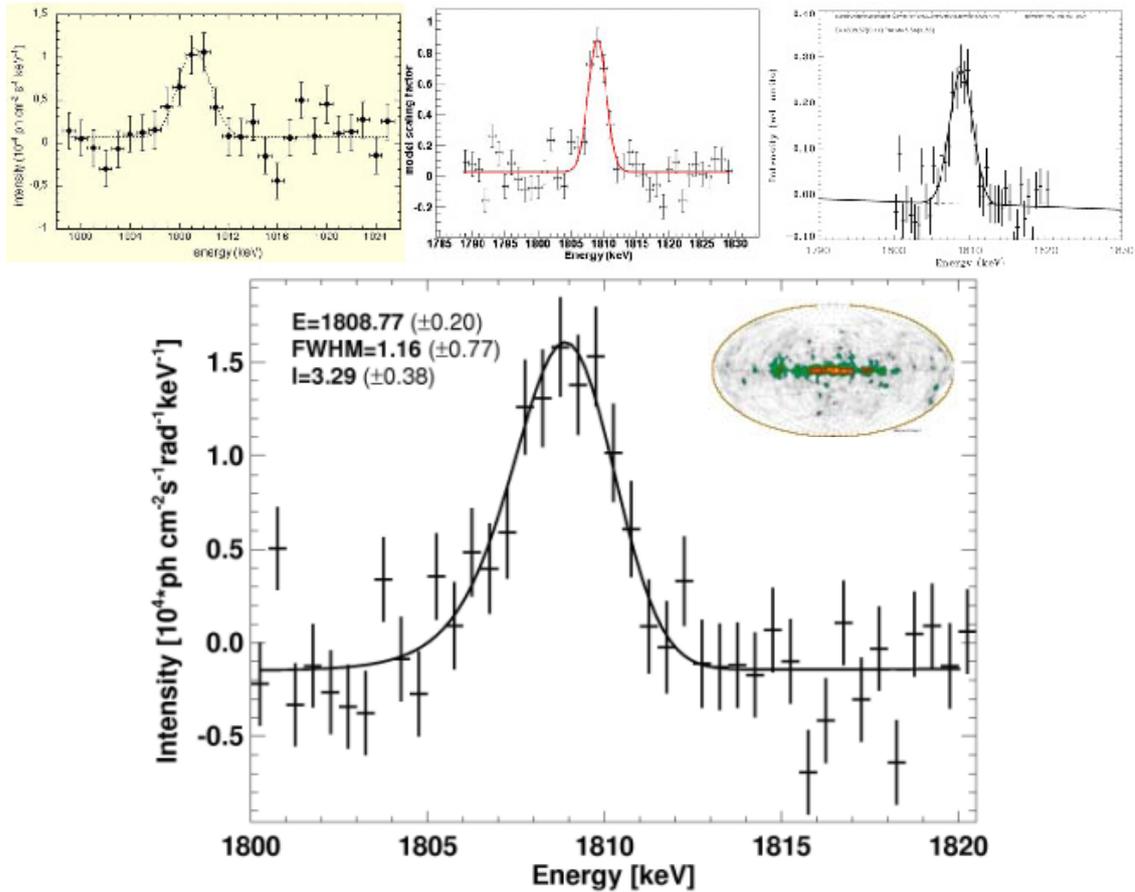

**FIGURE 3** $^{26}$Al line measurements of the SPI instrument: Early measurements [16] from 2003 and 2004 did not have adequate signal to constrain the line shape (top left and center figures); with 2 years of data [17] (top right and below), spectral precision appears sufficient for such investigations.

In this paper, we will discuss what has been learned from the first mission years of RHESSI and INTEGRAL with respect to nucleosynthesis sources in our Galaxy.

## $^{26}$AL LINE FLUX AND SPECTROSCOPY

From previous studies with the imaging Compton telescope instrument aboard the Compton Observatory (1991-2000 [8,9,10]), $^{26}$Al emission has been mapped all along the plane of the Galaxy. Conclusions from these imaging measurements were that massive stars dominate $^{26}$Al production, and that bright regions such as Cygnus suggest substantial $^{26}$Al ejection before the core collapse by massive stars in the Wolf-Rayet phase [11,12]. With respect to spectroscopy, the GRIS balloon experiment had reported[13] a line width of 6.4 keV, the celestial broadening of which would have corresponded to interstellar gas velocities of ~500 km s$^{-1}$. This was difficult to understand[14], further measurements of the line shape were important. Early results from RHESSI[15] and SPI[16] on INTEGRAL then showed that such a spectacularly-large line width was probably not real.

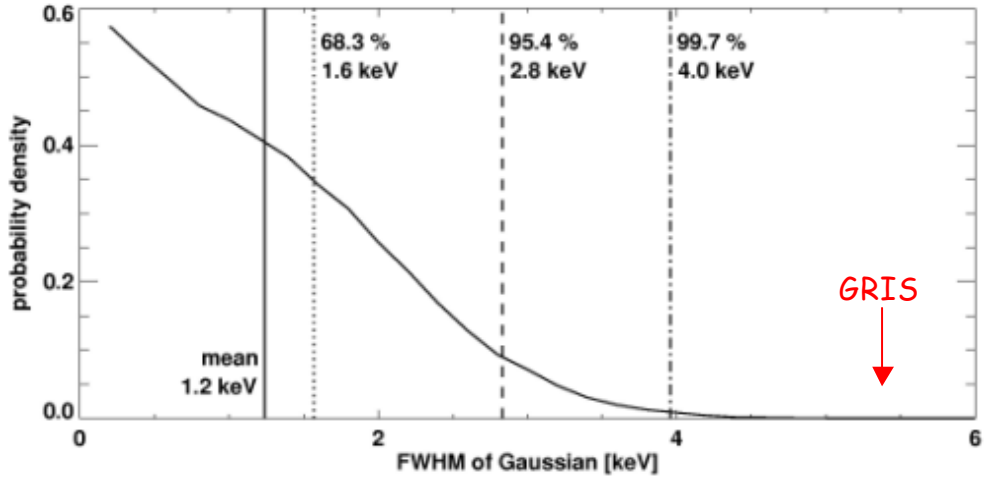

**Figure 4** Constraints on celestial broadening of the $^{26}$Al line[17]: The probability distribution is asymmetric, and values smaller than the formal mean of 1.2 keV are most probable. The GRIS value of 5.4 keV lies above 99.997% of all values.

With 2 years of INTEGRAL/SPI observations, the measurement of $^{26}$Al from the inner Galaxy is sufficient (~16σ, Fig.3) for analysis of the line shape details[17]. Modelling the shape with a convolution of the expected instrumental response and a Gaussian for a possibly broadened celestial line, it is found that any additional broadening from interstellar kinematics must be small. The quality of spectra was improved by a better determination of the SPI spectral response as it varies with time, maintained through periodic annealings against degradation of detectors (Fig.1). This now yields an upper limit (2σ) of 2.8 keV on celestial line broadening (Fig. 4), constraining the line width to more moderate and plausible values of ISM velocities (1 keV corresponds to 122 km s$^{-1}$) [17].

The integrated flux in the $^{26}$Al line obtained with INTEGRAL/SPI is 3.3±0.4 10$^{-4}$ ph cm$^{-2}$ s$^{-1}$; this is on the low side of previous measurements (see Figure 5).

From space-resolved spectroscopy, we could find now $^{26}$Al line shifts with Galactic longitude, which are consistent with expectations from differential Galactic rotation (Figure 6)[18]. This is a remarkable finding, because it allows us to conclude that the $^{26}$Al sources which we see towards the inner Galaxy indeed are populating the inner Galaxy as expected from candidate source distribution models. That means that the integrated flux from this region can be taken as a representative measurement of a source population throughout the Galaxy, and hence can be converted into a total Galactic $^{26}$Al mass produced by massive stars assuming a steady state (i.e. the star formation and supernova rates have remained constant over the past few mission years). We derive a Galactic amount of 2.8 ±0.8 M$_\odot$ of $^{26}$Al. This translates into a star formation rate of 3 ±1.4 M$_\odot$ yr$^{-1}$ or a core-collapse supernova rate of 1.9 ±1.1 events per century[18]. The global interstellar isotopic ratio $^{26}$Al/$^{27}$Al is 8.4 10$^{-6}$, comparing to a value for the early solar system of 4.5 10$^{-5}$. Those values are in agreement with a range of alternative methods, yet, unlike those, they are derived solely from measurements of our own Galaxy, and are free from major corrections for observational biases. The price paid in our case is a rather uncertain flux measurement due to the large instrumental background, and some dependency on the assumed Galactic source distribution model; both effects add to the substantial uncertainties we have to attach to our measurement.

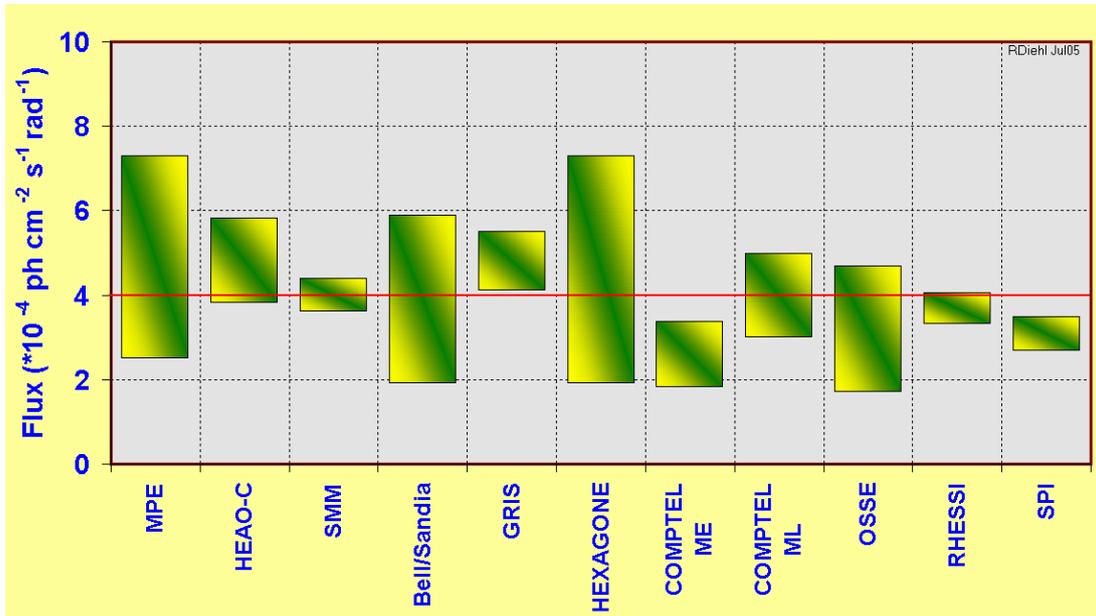

**FIGURE 5:** Comparison of the line fluxes for integrated $^{26}$Al emission from the inner Galaxy, normalized for the central radian.

Our recent imaging studies with SPI, from three years of data from a survey of the inner Galaxy and Cygnus region, indicate however a consistency of the emission mapping along the plane of the Galaxy, and hence support the conclusions obtained from COMPTEL results on large-scale source distributions [19]. We hope that the INTEGRAL mission will be extended into the next decade, and thus allow us to substantially improve on the present result through improved background determination.

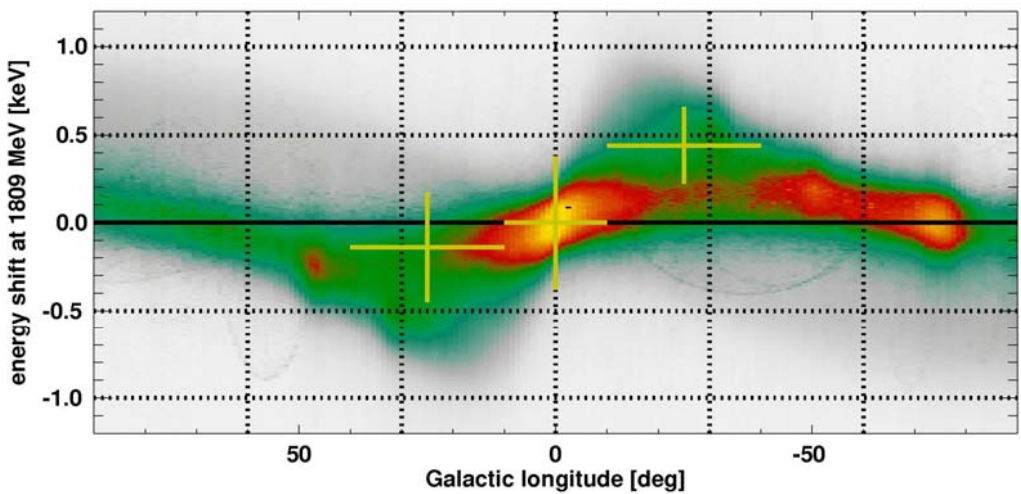

**FIGURE 6:** The space-resolved spectra of the $^{26}$Al line along the plane of the Galaxy show the expected signature from Doppler shifts of the line centroid due to Galactic rotation [18].

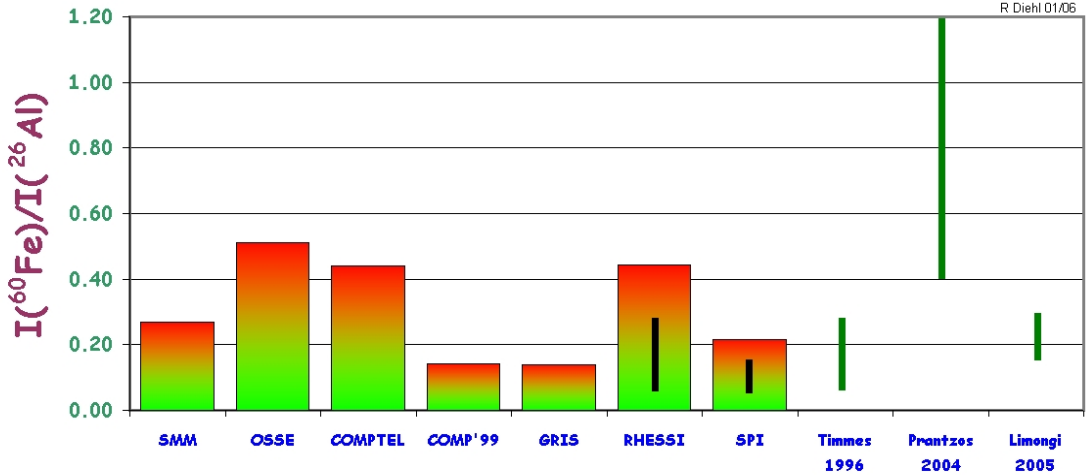

**FIGURE 7** Constraints on the $^{60}$Fe/$^{26}$Al gamma-ray intensity ratio from different gamma-ray measurements, and different evaluations of nucleosynthesis models, for comparison.

## $^{60}$FE PRODUCTION IN MASSIVE STARS

Both the RHESSI and SPI instrument have reported long-awaited detections of $^{60}$Fe decay gamma-rays from the inner Galaxy[20,21,22]. Both are detections at the 3σ level, hence may also be interpreted as upper limits; we show those, together with previous reported values, in Figure 7. Models of massive star nucleosynthesis have long predicted[23] that the massive stars producing $^{26}$Al should also be sources of $^{60}$Fe, and that the gamma-ray line intensities should be in the same order of magnitude. Predictions ten years ago showed a $^{60}$Fe/$^{26}$Al gamma-ray intensity ratio of 0.14 [23], while later studies showed that values above 1.0 seemed plausible[24]. Recent re-evaluation of models for the supernova yields into the interstellar medium from a population of massive stars appear in agreement (see Figure 7) with the recently-revised gamma-ray results[25]. But uncertainties are large both in theoretical models and in gamma-ray data, more homework is needed (Fig. 7).

## ACKNOWLEDGMENTS

This contribution results from the collaborative work with the members of the INTEGRAL and SPI Teams, and fruitful discussions with many other colleagues; I am grateful for their collaboration, in particular to Hubert Halloin, Karsten Kretschmar and Andrew Strong at MPE, Pierre Jean, Jürgen Knödlseder, and Jean-Pierre Roques at CESR, Trixi Wunderer at SSL Berkeley, Nikos Prantzos at IAP, Marco Limongi and Alessandro Chieffi at CNR Frascati, Dieter Hartmann at Clemson University, and Stan Woosley at UC Santa Cruz. INTEGRAL is an ESA project with instruments and science data centre funded by ESA member states (especially the PI countries: Denmark, France, Germany, Italy, Switzerland, Spain), Czech Republic and Poland, and with the participation of Russia and the USA. The SPI spectrometer has been completed under the responsibility and leadership of CNES/France, its anticoincidence system is supported by the German government through DLR grant 50.0G.9503.0. We acknowledge the support of INTEGRAL from ASI, CEA, CNES, DLR, ESA, INTA, NASA and OSTC.